\def\etal{{\frenchspacing\it et al.}}
\def\ie{{\frenchspacing\it i.e.}}
\def\eg{{\frenchspacing\it e.g.}}

\def\ith{i^{\rm th}}
\def\bth{b^{\rm th}}

\def\beq#1{\begin{equation}\label{#1}}
\def\eeq{\end{equation}}
\def\beqa#1{\begin{eqnarray}\label{#1}}
\def\eeqa{\end{eqnarray}}
\def\eq#1{equation~(\ref{#1})}
\def\Eq#1{Equation~(\ref{#1})}

\def\figsize{11.5cm}

\def\N{{\bf N}}
\def\e{{\bf e}}

\def\x{{\bf x}}

\def\Ni{{{\bf N}_b}}
\def\wi{{{\bf w}^b}}

\def\wt{{\tilde w}}
\def\xi{{{\bf x}^b}}

\def\expec#1{\langle#1\rangle}

\documentclass[twocolumn,aps,showpacs,showkeys,nofootinbib]{revtex4}

\newcommand{\be}{\begin{equation}}
\newcommand{\ee}{\end{equation}}
\newcommand{\ba}{\begin{eqnarray}}
\newcommand{\ea}{\end{eqnarray}}

\begin{document}
\input{epsf.sty}

\title{Uncorrelated Measurements of the Cosmic Expansion History and Dark Energy from Supernovae}
\author{Yun~Wang$^{1}$, and Max~Tegmark$^{2,3}$}
\address{$^1$ Department of Physics \& Astronomy, Univ. of Oklahoma,
                 440 W Brooks St., Norman, OK 73019;
                 email: wang@nhn.ou.edu}
\address{$^2$ Dept.~of Physics, Massachusetts Institute of Technology, Cambridge, MA 02139, USA; tegmark@mit.edu}
\address{$^3$ Dept.~of Physics, University of Pennsylvania, Philadelphia, PA 19104, USA}


\begin{abstract}

We present a method for measuring the cosmic expansion history $H(z)$ in uncorrelated redshift bins,
and apply it to current and simulated type Ia supernova data assuming spatial flatness.
If the matter density parameter $\Omega_m$ can be accurately measured from other data,
then the dark energy density history $X(z)=\rho_X(z)/\rho_X(0)$ can trivially be
derived from this expansion history $H(z)$.
In contrast to customary ``black box'' parameter fitting, 
our method is transparent and easy to interpret: the measurement of $H(z)^{-1}$ in a redshift bin
is simply a linear combination of the measured comoving distances for supernovae in that bin,
making it obvious how systematic errors propagate from input to output.

We find the Riess {\etal} (2004) ``gold'' sample to be consistent with the
``vanilla'' concordance model where the dark energy is a cosmological constant. 
We compare two mission concepts for the NASA/DOE Joint Dark Energy Mission (JDEM), 
the Joint Efficient Dark-energy Investigation (JEDI), and 
the Supernova Accelaration Probe (SNAP), using 
simulated data including 
the effect of weak lensing (based on numerical simulations)
and a systematic bias from K-corrections. 
Estimating $H(z)$ in seven uncorrelated redshift bins, we find that both provide dramatic improvements
over current data: JEDI can measure $H(z)$ to about 10\% accuracy and SNAP to 30-40\% accuracy.

\end{abstract}




\maketitle

\section{Introduction}

Observational data on type Ia supernovae (SNe Ia)
indicate that the expansion of our universe is
accelerating \citep{Riess98,Perl99}.
This can be explained by the presence of dark energy.
Various dark energy models have been considered, \eg$\,$ 
scalar fields
\citep{Freese87,Linde87,Peebles88,Wett88,Frieman95,Caldwell98}
and modified gravity \citep{sahni98,Parker99,Boisseau00,Deffayet01,Mersini01,Freese02,Carroll04}
--- see \cite{Pad} and \cite{Peebles03} for recent reviews.

Although current data seem consistent with a cosmological
constant (\eg,
\cite{Choudhury04,Chen04,Daly03,Daly04,Dicus04,Hannestad04,Simon04,WangTegmark,Jassal}), 
the uncertainties are large and more exotic models are not
ruled out (\eg, \cite{Alam04,Capozziello04,Huterer04,Feng04,Wang04a}). 
To uncover the nature of dark energy, and differentiate
among various dark energy models,
it is important that we extract dark energy constraints in
a model-independent manner \citep{WangGarnavich,WangLovelace,WangFreese}.
The perils of model assumptions and simplified parametrization of dark energy
have been shown in \cite{Maor02,WangTegmark,Bassett04}.

Throughout this paper, we assume spatial flatness as motivated by inflation.
Calibrated cosmological standard candles such as SNe Ia measure the luminosity
distance $d_L(z)= (1+z) r(z)$, where the comoving distance
\be\label{rEq}
r(z) = \int_0^z \frac{dz'}{H(z')} = H_*^{-1} \int_0^z \frac{dz'}{h(z')}
\ee
where $c=1$, $H_*=100$km$\,$s$^{-1}$Mpc$^{-1}$ and
\be\label{hEq}
h(z)\equiv H(z)/H_* =h(0)\left[ \Omega_m (1+z)^3 + \Omega_X X(z)\right]^{1/2},
\ee
with $X(z)\equiv \rho_X(z)/\rho_X(0)$ denoting the dark energy density
function. 
Determining if and (if so) how the dark energy density $X(z)$ 
depends on cosmic time is the main observational goal in the current quest to
illuminate the nature of dark energy.
Given a precise measurement of
the matter density fraction $\Omega_m$ (from galaxy redshift surveys, for example),
the dark energy density function $X(z)$ 
can be trivially determined from $H(z)$ via \eq{hEq}.

Numerous methods for this have been developed and applied in the recent literature,
either by parametrizing $X(z)$ in terms of an equation of state and perhaps
additional parameters or by aiming for more model-independent constraints
(\eg, \cite{WangTegmark}).
\Eq{rEq} shows that the data directly constrain the cosmic expansion history
$H(z)$ --- in principle. The information-theoretically minimal error bars on $H(z)$ that can
be obtained from supernovae were derived in \cite{gravity} using a Fisher matrix approach.
Yet no optimal method for doing this in practice has been found other than the ``black box'' 
approach of parametrizing $H(z)$ somehow and fitting to the data. 
Ideally one would like to measure $H(z)$ in many redshift bins with uncorrelated error
bars, but \cite{gravity} found that parametrized fits tend to yield 
broad and difficult-to-interpret window functions, \ie, the measurement in a given redshift bin
depended also on supernova data far outside that redshift range.
\cite{Huterer04} strengthened this conclusion by showing that uncorrelated 
measurements of the expansion history (computed by diagonalizing the Fisher matrix) tended to 
probe a broad redshift range. The approach most similar to ours is that of \cite{gravity,Daly03,Daly04},
where $H(z)^{-1}$ is measured by numerically differentiating \eq{rEq}. These papers tackle the 
challenge of differentiating sparse and noisy data by performing a polynomial fit for $r(z)$:
\cite{gravity} performs a global fit whereas \cite{Daly03,Daly04} improve this with a local fit within a 
sliding window.

The purpose of the present paper is to solve this problem optimally, 
presenting a method giving uncorrelated measurements of the expansion history in 
arbitrary redshift bins. We will see that this method is both 
easy to implement and easy to interpret.
We describe our method in Sec.2.   
We present our results in Sec.3 and discuss our conclusions in Sec.4.


\begin{figure} 
\epsfxsize=\figsize\epsffile{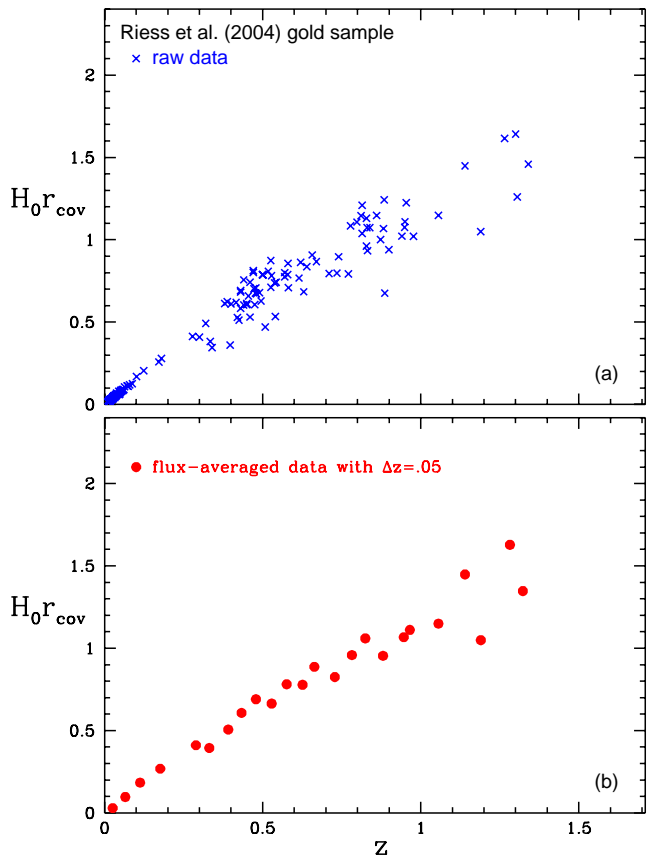}
\caption[1]{\label{rFig}\footnotesize%
Data used. 
The measured dimensionless comoving distances are shown for the Riess {\etal} (2004) ``gold'' sample 
before (top panel) and after (bottom panel) flux averaging. 
The cosmic expansion $h(z)$ is simply the inverse of the derivative of this, so the visually obvious 
fact that the derivative is falling immediately tells us that the universe was denser and faster expanding in the past.
}
\end{figure}

\section{Method}

Assuming that the redshifts of SNe Ia are accurately measured,
we can neglect redshift uncertainties, and simply treat the measured 
comoving distances $r(z_i)$ to the supernovae (plotted in Figure 1) as the observables.
In terms of $\mu_0$, the distance modulus of SNe Ia, we have
\be
\frac{r(z)}{\mbox{1 Mpc}}=\frac{1}{2997.9(1+z)}
\, 10^{\mu_0/5-5}.
\ee
Let us write the comoving distance measured from the $\ith$
SN Ia as
\be
r_i = r(z_i) + n_i,
\ee
where the noise vector satisfies $\expec{n_i}=0$, $\expec{n_i n_j} = \sigma_i^2 \delta_{ij}$.

\subsection{Transforming to measurements of $H(z)^{-1}$}

As a first step in our method, 
we sort the supernovae by increasing redshift $z_1<z_2<...$, and define the quantities
\ba\label{xEq}
x_i & \equiv & \frac{r_{i+1}-r_i}{z_{i+1}-z_i}
= \frac{ \int_{z_i}^{z_{i+1}} \frac{dz'}{H(z')} + n_{i+1}-n_{i}}
{z_{i+1}-z_i} \nonumber \\
& = & \overline{f}_i + (n_{i+1}-n_{i})/\Delta z_i,
\ea
where $\Delta z_i \equiv z_{i+1}-z_i$ and
$\overline{f}_i$ is the average of $1/H(z)$ over the
redshift range $(z_i, z_{i+1})$.
Note that $x_i$ gives an unbiased estimate of the average of $1/H(z)$ in the redshift bin, since
$\langle x_i \rangle = \overline{f}_i$, so the quantities $x_i$ are direct (but noisy) probes
of the cosmic expansion history.
Assembling the numbers $x_i$ into a vector $\x$, 
its covariance matrix $\N\equiv\expec{\x\x^t}-\expec{\x}\expec{\x}^t$ is tridiagonal, 
satisfying $N_{ij}=0$ except for the following cases:
\ba
N_{i,i-1}&=& -\frac{\sigma_i^2}{\Delta z_{i-1} \Delta z_i},\nonumber\\
N_{i,i}  &=& \frac{\sigma_i^2+ \sigma_{i+1}^2}{\Delta z_i^2}, \nonumber\\
N_{i,i+1}&=& -\frac{\sigma_{i+1}^2}{\Delta z_i \Delta z_{i+1}}.
\ea
The new data vector $\x$ clearly retains all the cosmological information from the original data set
(the comoving distance measurements $r_i$), since the latter can trivially be recovered from $\x$
up to an overall constant offset by inverting \eq{xEq}.
In summary, the transformed data vector $\x$ expresses the SN Ia information as 
a large number of unbiased but noisy measurements of the cosmic expansion history in
very fine redshift bins, corresponding to the redshift separations between neighboring supernovae.

\begin{figure} 
\epsfxsize=\figsize\epsffile{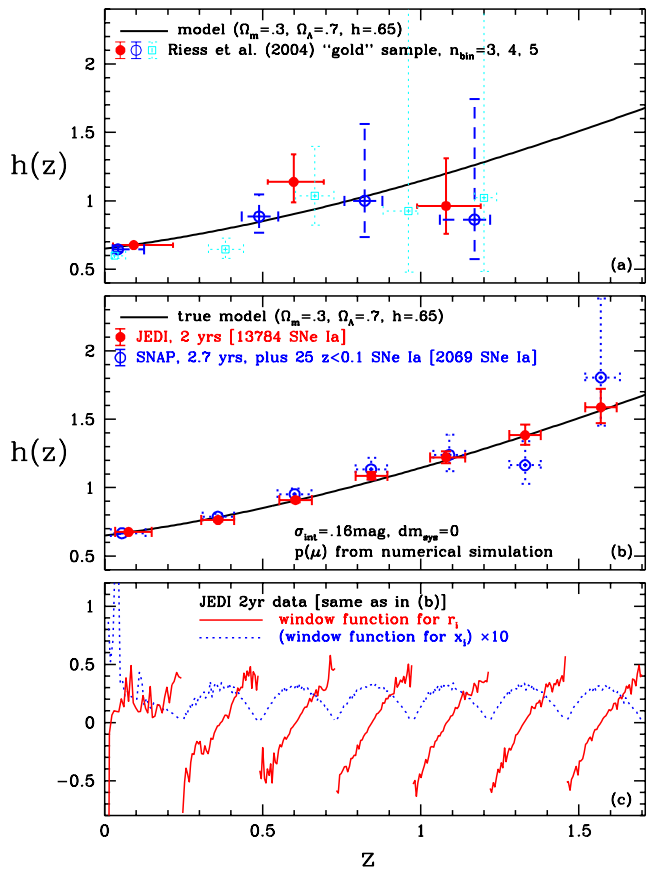}
\caption[1]{\label{hFig}\footnotesize%
The cosmic expansion history (the dimensionless Hubble parameter $h(z)$) 
is measured in uncorrelated redshift bins from the Riess {\etal} (2004) ``gold'' sample (top panel)
and from simulated future data (middle panel) for 
the NASA/JDEM mission concepts JEDI (solid points) and SNAP (dotted points). 
The measured $h(z)^{-1}$ in a given redshift bin is simply the sum of
the comoving supernova distances in that bin, weighted by the corresponding solid curve
in the bottom panel, which roughly speaking subtracts more nearby supernovae from more distant ones.
}
\end{figure}

\subsection{Averaging in redshift bins}

The second step in our method is to average these noisy measurements $\x$ 
into minimum-variance measurements $y_b$ of the expansion history in some given redshift bins.
For instance, the middle panel of Figure 2 shows an example with seven bins, $b=1,...,7$.
Let the vector $\xi$ denote the piece of the $\x$-vector corresponding to the $\bth$ bin,
and let $\Ni$ denote the corresponding covariance matrix.
Our weighted average $y_b$ can then be written 
\be
y_b=\wi\cdot\xi
\ee
for some weight vector $\wi$ whose components sum to unity, 
\ie, $\sum_i w^b_i=1$ or equivalently 
$\e\cdot\wi =1$, where $\e$ is a vector containing all ones; $e_i=1$.
To find the best weight vector $\wi$, we
minimize the variance
\be\label{DyEq}
\Delta y_b^2\equiv\expec{y_b^2}-\expec{y_b}^2=\wi^t\Ni\wi
\ee
subject to the constraint that the weights add up to unity, \ie, that $\e\cdot\wi=1$.
This constrained minimization problem is readily solved with the Lagrange multiplier method,
giving 
\be
\wi = \frac{\Ni^{-1}\e}{\e^t\Ni^{-1}\e},
\ee
and substituting this back into \eq{DyEq} gives the size of the corresponding error bar:
\be
\Delta y_b = \left(\e^t\Ni^{-1}\e\right)^{-1/2}.
\ee
The bottom panel in Figure 2 shows the seven weight vectors
$\wi$ (dotted) corresponding to the seven measurements $y_b$ in the 
middle panel (solid points). We will also refer 
to the weight vectors as {\it window functions},
since they show the contributions to our measurements from different redshifts.
Note that each window function vanishes outside its
redshift bin, and that all seven of them share a characteristic
bump shape roughly corresponding to  
an upside-down parabola vanishing at the bin endpoints.
To illustrate the $z$-range that each measurement probes, we
plot it at the median of the window function with horizontal bars
ranging from the 20th to the 80th percentile.

Since the measurements $y_b$ are linear combinations of the $x_i$
which are in turn linear combinations of the $r_i$,
we can also reexpress our measurements directly as linear combinations
of the original supernova comoving distances:
\be\label{wpEq1}
y_b=\sum_i\wt^b_i r_i,
\ee
where the new window functions
\be\label{wpEq2}
\wt^b_i\equiv {w^b_{i-1}\over\Delta z_{i-1}} - {w^b_i\over\Delta z_i}
\ee
would be essentially the negative derivative of the old window functions if all redshift intervals
were the same. 
These new window functions are also plotted in the bottom panel of Figure 2 (solid ``sawtooth'' curves),
and are seen to be roughly linear (as expected for a parabola derivative), effectively subtracting
supernovae at the near end of the bin from those at the far end.

A major advantage of this method is its transparency and simplicity. 
If one fits some parametrized model of $H(z)$ to the SN Ia data by maximizing a likelihood function, 
then the resulting parameter estimates will be some complicated (and generically nonlinear) functions
of all the data points $r_i$.
In contrast, the measurement $y_5$ in our $5^{\rm th}$ bin in figure 2 (middle panel) is 
simply a linear combination of the comoving distance measurements for the supernovae in 
the $5^{\rm th}$ bin ($1.0<z_i<1.4$) as defined by the $5^{\rm th}$ window function 
in the bottom panel, so it is completely clear how each particular supernova affects the final result.
In particular, the supernovae outside of this redshift range do not affect the measurement at all.

We conclude this section by discussing some details useful for the reader interested in applying
our method in practice.

\subsection{Creating uncorrelated redshift bins}

We suggest discarding those $x_i$ straddling neighboring bins, \ie, whose
two supernovae fall on either side of a bin boundary. For say 7 bins there are only 6 such numbers, so
this involves a rather negligible loss. The advantage is that it ensures that two measurements 
$y_b$ and $y_{b'}$ have completely uncorrelated error bars if $b\ne b'$, since their window functions $\wt$
have no supernovae in common.

\subsection{Flux averaging}

To minimize the bias in the $H(z)$-measurement due to weak lensing, we
use flux-averaging \citep{Wang00b,WangMukherjee}\footnote{A Fortran 
code that uses flux-averaging statistics 
to compute the likelihood of an arbitrary 
dark energy model (given the SN Ia data from \cite{Riess04})  
can be found at $http://www.nhn.ou.edu/\sim wang/SNcode/$.}.
Specifically, we compress the full supernova data set $r_i$ into a smaller number 
of of flux-averaged supernovae assigned to the mean redshift in each of a large number of
bins of width $\Delta z$. We use $\Delta z=0.05$ for the current data and 
$\Delta z=0.005$ for the simulated data.
Note that since we have assumed a Gaussian distribution in the magnitudes
of SNe Ia at peak brightness, flux-averaging leads to a tiny
bias of $- \sigma_{int}^2 \, \ln 10 /5$ mag \citep{Wang00a}.
We have removed this tiny bias in the data analysis.

As a side effect, this averaging in narrow bins makes the 
denominators $\Delta z_i$ roughly equal in \eq{xEq},
so that the only noticeable source of wiggles in the window functions in Figure 2 (bottom)
is Poisson noise, \ie, that some of these narrow bins contain more supernovae than others.
The method of course works without this averaging step as well.
In that case, the window functions ${\bf w}$ wiggle substantially because of 
variations in the redshift spacing between supernovae, since very little 
weight is given to $x_i$ if $\Delta z_i$ happens to be tiny. 
However, we find that the supernova window functions $\wt$ remain rather 
smooth and well-behaved functions of
redshift, as expected --- two supernovae very close together with the same noise 
level $\sigma_i$ automatically get the same weight.
This means that our method effectively averages such similar 
redshift supernovae anyway, even if we do
not do so by hand ahead of time. The difference between flux 
averaging and this automatic averaging
is simply that we average their fluxes rather than their comoving 
distances --- these two types of averaging are not equivalent
since the flux is a nonlinear function of the comoving distance.

\section{Results}

Figure 2 shows the results of applying our method to both real data (top panel) 
and simulated data (middle panel),
with the dimensionless expansion rate of the Universe $h(z)$ 
measured in between three and seven uncorrelated redshift bins.

The top panel uses the ``gold'' set of 157 SNe Ia published by \cite{Riess04}.
The error bars are seen to be rather large, and consistent with a simple flat 
$\Omega_m=0.3$ concordance model
where the dark energy is a cosmological constant.\footnote{This is consistent
with the findings of \cite{WangTegmark}.}
As was shown in \cite{gravity} using information theory, the relative error 
bars on the cosmic expansion history $H(z)$ 
scale as 
\be\label{ErrorEq}
{\Delta H\over H} \propto {\sigma\over N^{1/2}\Delta z^{3/2}},
\ee
for $N$ supernovae with noise $\sigma$.
Here $\Delta z$ is the width of the redshift bins used, so one pays a great price for narrower 
bins: halving the bin size requires eight times as many supernovae.
The origin of this $(\Delta z)^{-3/2}$-scaling is intuitively clear: the noise 
averages down as $(\Delta z)^{-1/2}$, and there is
an additional factor of $(\Delta z)^{-1}$ from effectively taking the derivative of the data 
to recover $H(z)^{-1}$ from the integral in \eq{xEq}\footnote{Anaogous estimates 
of the equation of state $w(z)$
have a painful $(\Delta z)^{-5/2}$-scaling, since they effectively involve taking 
the second derivative of the data.}.
The bottom panel in Figure 2 shows that the method effectively estimates this 
derivative by subtracting supernovae at the
near end of the bin from those at the far end of the bin and dividing by $\Delta z$.

\begin{figure} 
\vskip-6.9cm
\epsfxsize=\figsize\epsffile{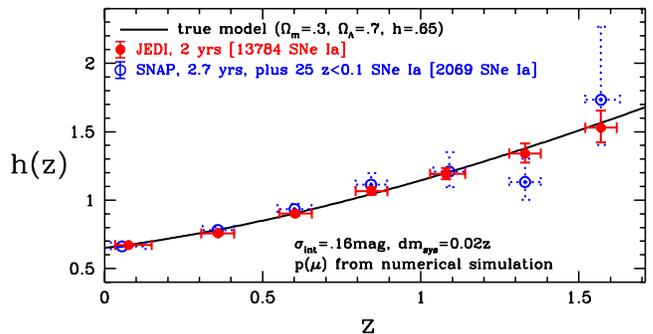}
\caption[1]{\label{hFig2}\footnotesize%
Same as Fig.2(b), but with a systematic bias of $dm_{sys}=0.02\,z$ from 
K-correction uncertainties added in addition to lensing noise computed 
from numerical simulations (Barber 2003 \cite{Barber}).
}
\end{figure}

This means that for accurately measuring $H(z)$ and thereby the density history of the dark energy,
numbers really do matter. For example, 
in order to measure the dark energy density function to 10\% accuracy
in seven uncorrelated redshift bins as in the middle panel of Figure 2 (with $\Delta z=0.243$), 
we need to have around 14,000 SNe Ia.
We have simulated SN Ia data by placing supernovae
at random redshifts, with the number of SNe Ia per 0.1
redshift interval given by a distribution.
The intrinsic brightness of each SN Ia at peak brightness 
is drawn from a Gaussian distribution with a dispersion of
$\sigma_{int}=0.16\,$ mag.
As the fiducial cosmological model, we used a flat universe with $\Omega_{\Lambda}=0.7$.

We compare two mission concepts for the NASA/DOE Joint Dark 
Energy Mission (JDEM): the 
Joint Efficient Dark-energy Investigation (JEDI) \citep{Wang04b} and the 
Supernova Accelaration Probe (SNAP) \citep{SNAP}.
For JEDI (solid points in Figure 2), the number of SNe Ia per 0.1
redshift interval is obtained by fitting the measured 
SN Ia rate as function of redshift \citep{Cappellaro99,Hardin00,Dahlen04}
to a model assuming
a conservative delay time between star formation and
SN Ia explosion of 3.5 Gyrs.
For SNAP (dotted points in Figure 2), the number of SNe Ia per 0.1
redshift interval is taken from Figure 9 in \cite{SNAP}.

We consider two kinds of SN Ia systematic uncertainties:
weak lensing due to intervening matter and a systematic
bias due to K-corrections.
We include the weak lensing effect by assigning a magnification
$\mu$ drawn from a probability distribution $p(\mu)$, extracted
using an improved version of the Universal Probability Distribution 
Function (UPDF) method \citep{Wang02}
from the numerical simulations of weak lensing
by Barber (2003) \cite{Barber}.
The total uncertainty in each SN Ia data point is
$\sqrt{\sigma_{int}^2 + \sigma_{lens}(z)^2}$, with
$\sigma_{lens}(z)$ extracted from Barber (2003) \cite{Barber}:
\be
\sigma_{lens}(z) \simeq 
0.00311 + 0.08687 z-0.00950 z^2
\ee
We consider a systematic bias of $\Delta m_{\rm sys}=0.02 z$
due to K-corrections following \cite{WangGarnavich}.

We did not include the systematic bias due to K-corrections in
Fig.2, in order to compare the real data \citep{Riess04} and
simulated data on an equal footing.

In {Fig.}~3, we show the effect of adding the systematic bias
due to K-corrections in addition to the weak lensing noise.
Comparing {Fig.}~3 with {Fig.}~2 (b), we see that the systematic bias
does not have a significant effect on the uncorrelated
estimates of $H(z)$. This is because our method effectively
reduces a global systematic bias into a local bias with
a much smaller amplitude (see {Sec.}~2).

\begin{figure} 
\epsfxsize=\figsize\epsffile{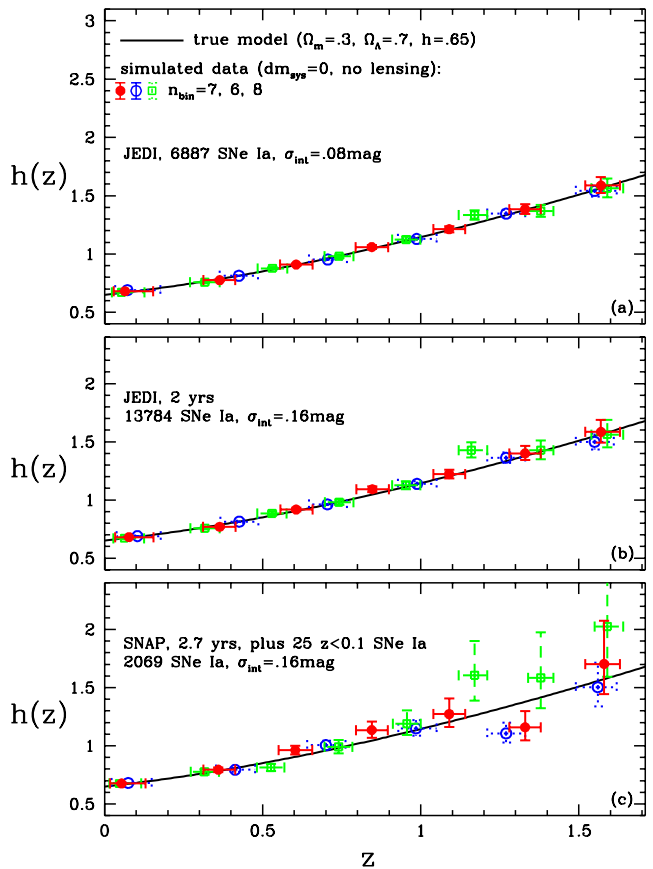}
\caption[1]{\label{hFig3}\footnotesize%
How the recovery of the cosmic expansion history $h(z)$ depends on
the number of redshift bins, 
assuming no systematic bias and no lensing.
(a) Half of JEDI data, with a reduced intrinsic scatter of
$\sigma_{int}=0.08\,$ mag.
(b) All of JEDI data, with $\sigma_{int}=0.16\,$ mag.
(c) All of SNAP data (plus 25 SNe Ia at $z<0.1$), with $\sigma_{int}=0.16\,$ mag.
}
\end{figure}

Figure 4 shows how the recovery of the cosmic expansion history $h(z)$ depends on
the number of redshift bins (6, 7 and 8), assuming no systematic bias and no lensing, and agrees
well with the theoretical $(\Delta z)^{3/2}$ scaling.
Contrasting current and future data with roughly the same 
redshift binsize for both ($n_{bin}$=5 for current data,
and $n_{bin}$=7 for the simulated data) shows that 
JEDI shrinks the error bars by more than an order of magnitude,
so the potential improvement with a successful JDEM would be dramatic.

Figure 4 compares three different data sets:
(a) Half of the JEDI data, with a reduced intrinsic scatter of
$\sigma_{int}=0.08\,$ mag from sub-typing.
(b) All the JEDI data, with $\sigma_{int}=0.16\,$ mag.
(c) All the SNAP data (plus 25 SNe Ia at $z<0.1$), with $\sigma_{int}=0.16\,$ mag.

Note that since the measured quantity $h(z)^{-1}$ typically 
is not a straight line, the measured average of this curve over a 
redshift bin will generally lie either slightly above of below 
the curve at the bin center. Figure 2 shows that 
this bias is substantially smaller 
than the measurement uncertainties, since $h(z)$ and $h(z)^{-1}$ are
rather well  approximated by straight lines over the narrow redshift bins
that we have used.

A second caveat when interpreting our figures is that the absolute calibration 
of SN Ia is not perfectly known --- changing this simply corresponds to multiplying the function 
$h(z)$ by a constant, \ie, to scaling the measured curve vertically.

\section{Discussion}

We have presented a method for measuring the cosmic expansion history $H(z)$ in uncorrelated redshift bins,
and applied it to current and simulated supernova Ia data assuming spatial flatness.
Whereas previously proposed approaches involve ``black box'' parameter fitting, 
this method is transparent and simple to interpret: the measurement of $H(z)^{-1}$ in a redshift bin
is simply a linear combination of the measured comoving distances for supernovae in that bin, with
weights that roughly correspond to subtracting closer supernovae from more distant ones.
Such transparency is particularly helpful for understanding how systematic errors in the input affect the output.
For instance, a constant systematic dimming throughout a redshift bin will leave that measurement unaffected.

This method is useful for understanding the nature of dark energy, 
since the dark energy density history follows from 
this expansion history $H(z)$ if 
the matter density parameter $\Omega_m$ can be accurately measured from other data 
(\eg, the cosmic microwave background, galaxy clustering and gravitational lensing).
We found the Riess {\etal} (2004) ``gold'' sample to be consistent with the
``vanilla'' concordance model where the dark energy is a cosmological constant, but that much larger numbers
of supernovae are needed for true precision tests of the nature of dark energy.
We obtain good agreement with the results of \cite{Daly03,Daly04} --- the fact that two quite different techniques
give consistent answers indicates a reassuring robustness to method and data details.

Looking towards the future, we compare two mission concepts for the NASA/DOE Joint Dark Energy Mission (JDEM), 
the Joint Efficient Dark-energy Investigation (JEDI) and 
the Supernova Accelaration Probe (SNAP), using 
simulated data including 
the effect of weak lensing and bias from K-corrections. 
Estimating $H(z)$ in seven uncorrelated redshift bins, we find that both provide dramatic improvements over
the present state-of-the art: JEDI can measure $H(z)$ to about 10\% accuracy, and SNAP can measure
$H(z)$ to 30-40\% accuracy (Figure 2).

Our results show that numbers do matter. For example, 
in order to measure the cosmic expansion history to 10\% accuracy
in seven uncorrelated redshift bins (with $\Delta z=0.243$), 
we need to have around 14,000 SNe Ia
(as expected from two years of JEDI data), assuming
a dispersion of 0.16 magnitudes in SN Ia peak brightness (Figure 2).
Estimating $H(z)$ to 10\% in smaller uncorrelated redshift bins 
will require an even larger number of SNe Ia (if a significant 
reduction in intrinsic dispersion is not assumed for a sizable
fraction of the SNe Ia), which will be difficult to obtain in
a feasible two-year space mission. Also, $H(z)$ estimated in
$\Delta z=0.243$ redshift bins (for $0\leq z \leq 1.7$)
to 10\% accuracy will give us 
a powerful means to differentiate between a cosmological constant
and dark energy models which are not fine-tuned to mimic a
cosmological constant. 
A sample with a large number of SNe Ia allows tighter calibration of
SNe Ia as standard candles and subtyping of SNe Ia to reduce
diversity. This may yield a smaller set of SNe Ia with
substantially smaller intrinsic dispersion, which can lead
to more robust and stable estimates of the expansion history
of the universe (Figure 4a).
If the subtyping works well in reducing the
intrinsic dispersion of SNe Ia, we can expect to be
able to measure $H(z)$ in more than seven redshift bins
to 10\% accuracy for $0\leq z \leq 1.7$ (Figure 4a).

A precise measurement of the cosmic expansion history 
as a free function of cosmic time to 10\% accuracy would represent a 
dramatic improvement in our knowledge about dark energy.
Our results suggest that a JDEM can achieve this scientific goal.

\bigskip

{\bf Acknowledgements}
We thank David Branch, Peter Garnavich, and Ruth Daly for helpful comments.
This work was supported by
NSF CAREER grants AST-0094335 (YW) and AST-0134999 (MT), 
NASA grant NAG5-11099 and fellowships from the David and Lucile
Packard Foundation and the Cottrell Foundation (MT).

\end{document}